\documentstyle[12pt,epsfig,psfig]{article}

\newcommand{\ds}{\displaystyle}

\begin{document}

\noindent December 1998 \hfill BI-TP 98/36

\bigskip~
\bigskip

\begin{center}
{\large\bf $J/\psi$-Photoproduction\\[0.2cm]
and the Gluon Structure of the Nucleon}\\[0.5cm]

\medskip

{D.\ Kharzeev$^1$, H.\ Satz$^2$, A.\ Syamtomov$^3$ and
G.\ Zinovjev$^2$$^,$$^3$}\\[0.5cm]

\medskip

{$\mbox{}^1$ Riken/BNL Institute, Brookhaven National Laboratory\\
Upton/NY 11973, USA\\[0.2cm]
$\mbox{}^2$Fakult\"{a}t f\"{u}r Physik,
Universit\"{a}t Bielefeld,\\ D-33501 Bielefeld, Germany\\[0.2cm]
$\mbox{}^3$Bogolyubov Institute for Theoretical Physics,\\
National Academy of Sciences of Ukraine,\\
UA-252143 Kiev, Ukraine}
\end{center}

\bigskip

\centerline{\bf Abstract:}

\bigskip

Using short distance QCD methods based on the operator product expansion,
we calculate the $J/\psi$ photoproduction cross section in terms of the gluon
distribution function of the nucleon. Comparing the result with data,
we show that experimental behaviour of the cross section correctly 
reflects the $x$-dependence of the gluon distribution obtained from
deep inelastic scattering.

\vskip 1cm

The charmonium ground state $J/\psi$ is much smaller than the conventional
hadrons constructed from light $u$ and $d$ quarks, and much more tightly bound: 
we recall that $r_{J/\psi} \simeq 0.2\ {\rm fm} \ll \Lambda^{-1}_{QCD}$ and
$2M_D - M_{J/\psi} \simeq 0.64\ {\rm GeV} \gg \Lambda_{QCD}$. As a result, when 
a $J/\psi$ interacts with a `light' hadron, it is expected to probe the local
partonic structure of the latter, not its size or mass. The aim of this
paper is to show that measurements of $J/\psi$-photoproduction on nucleons  
confirm this expectation and give a rather precise reflection of the 
gluon distribution function of a nucleon.

\medskip

Using vector meson dominance (VMD), we shall first relate forward 
$J/\psi$-photo\-pro\-duction to $J/\psi$-nucleon scattering. Next, the total
$J/\psi$-nucleon cross section is expressed in terms of nucleonic gluon 
distribution functions, making use of the short distance QCD methods based 
on sum rules derived from the operator product expansion (OPE). The last missing 
element, the ratio of real and imaginary $J/\psi - N$ scattering amplitudes, 
is obtained by a dispersion relation. We can then study directly the 
interrelation of the energy dependence of $J/\psi$-photoproduction and the 
$x$ dependence of the gluon distribution in the nucleon.

\medskip

The use of the $J/\psi$ as a probe of the partonic status of a given medium
is of particular interest for the study of colour deconfinement in
nucleus-nucleus collisions. An essential and here particularly relevant 
aspect of deconfinement is that the constituent partons of a deconfined
medium are 
no longer constrained to the distribution functions of individual hadrons, 
as determined in deep inelastic scattering. The hadronic gluon distribution 
functions are strongly suppressed at high gluon momenta, with 
$xg(x) \sim (1-x)^a$ for $x \rightarrow 1$, where $a \simeq 3-5$ and 
$x=k_{g}/k_{h}$ is the relative fraction of the hadron momentum carried by 
the gluon. Removing this constraint will generally lead to harder gluons. 
Since $J/\psi$ dissociation requires hard gluons, the inelastic 
$J/\psi$-hadron cross section becomes very small at low collision energies. 
Hence significant $J/\psi$ suppression in nuclear collisions requires
the produced environment to be deconfined.

\medskip

In principle, the predicted threshold suppression of the $J/\psi$-hadron 
dissociation cross section can be measured directly. However, until such 
experiments are carried out, and to check the universality of the phenomenon, 
it is of interest to consider other processes in which the partonic aspects
of the $J/\psi$-hadron interactions play a decisive role. The photoproduction
of heavy quarkonium states is, as we shall see, an excellent case at hand, 
with a quite extensive set of experimental data available over a wide region 
of incident energies.

\medskip

Within the conventional VMD approach, we can relate the reactions 
$\gamma N \rightarrow \psi N$ and $\psi N \rightarrow \psi N$ and express 
the differential cross section of the forward $J/\psi$-photoproduction 
on nucleons as \cite{BargerPhillips} 
\begin{equation}
\frac{d\,\sigma_{\gamma\,N\rightarrow \psi\,N}}{d\,t}(s,t=0) =
\frac{3\Gamma(\psi\rightarrow e^+e^-)}{\alpha m_{\psi}}
\left(k_{\psi N}\over k_{\gamma N}\right)^2
\frac{d\,\sigma_{\psi\,N\rightarrow \psi\,N}}{d\,t}(s,t=0)
\label{VMD0}
\end{equation}
where $k^{2}_{ab}=[s-(m_{a}+m_{b})^2][s-(m_{a}-m_{b})^2]/4s$ denotes the
squared center of mass momentum of the corresponding reaction and $\Gamma$
stands for the partial $J/\psi$-decay width; to simplify
formulae, we shall in most expressions abbreviate $J/\psi$ by $\psi$. The
differential $J/\psi - N$ cross section is given by
\begin{equation}
\frac{d\,\sigma_{\psi\,N\rightarrow \psi\,N}}{d\,t}(s,t=0) =
\frac 1 {64\pi} \frac 1 {m_{\psi}^2(\lambda^2-m_N^2)}  
|{\cal M}_{\psi\,N} (s,t=0)|^2 
\label{VMD00}
\end{equation}
with $\cal M_{\psi\,N}$ denoting the invariant $J/\psi - N$ scattering 
amplitude; further, $\lambda =(pK/m_{\psi})$ is the nucleon energy in the 
quarkonium rest frame and $p,K,q$ are the four-momenta of target nucleon, 
$J/\psi$ and initial photon, respectively. From the optical theorem in the form
\begin{equation}
\sigma_{\psi\,N}^{tot}=\frac{Im\ {\cal M}_{\psi\,N}}{2m_{\psi}
\sqrt{\lambda^2-m_N^2}},
\label{OptTheor}
\end{equation}
we then obtain the well-known relation
\begin{equation}
\frac{d\,\sigma_{\gamma\,N\rightarrow \psi\,N}}{d\,t}(s,t=0)=
\frac{3\Gamma(\psi\rightarrow
e^+e^-)[s-(m_{N}+m_{\psi})^2][s-(m_{N}-m_{\psi})^2](1+\rho^2)}{16\pi\alpha~
m_{\psi}(s-{m_N}^2)^{2}}
\left({\sigma}_{\psi\,N}^{tot}\right)^2,
\label{VMD1}
\end{equation}
where $\rho$ is the ratio of real and imaginary parts of forward $J/\psi-N$
scattering amplitude.

\begin{figure}[htb]
\vspace*{-0mm}
\epsfysize=10cm
\centerline{\epsfig{file=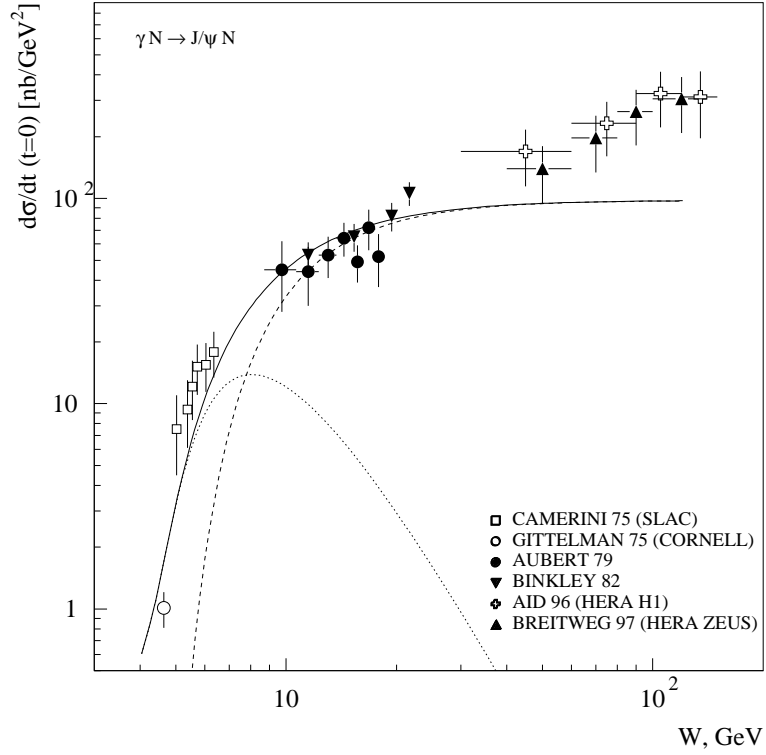,width=10cm}}
\caption{Forward $J/\psi$ photoproduction
data compared to our results with (solid line) and without (dashed line)
the real part of the amplitude.  
The curves were obtained using a scaling PDF \cite{KS94}}
\end{figure}

By use of the operator product expansion \cite{Peskin}-\cite{KSSZ},
the $J/\psi-N$ scattering amplitude in the unphysical region $\lambda\sim 0$
is obtained in terms of the nucleonic gluon distribution $g(x,\epsilon_0^2)$
\begin{eqnarray}
& & {\cal M}_{\psi N} = 2\,m_{\psi}\sqrt{\pi}r_0^3\epsilon_0^2
\left({32\over3}\right)^2\left[\int_0^1\, dx
\sum_{n=2,4,\ldots}^\infty x^{n-2}
\left({{\lambda}\over{\epsilon_{0}}}\right)^n
\,g(x,\epsilon_0^{2})\times\right.\nonumber\\
& & \times {{\Gamma(n+{5\over2})}\over{\Gamma(n+5)}}
\,{}_3F_2\left({5\over4}+{n\over2},{7\over4}+{n\over2},1+n;
{{(5+n)}\over{2}},3+{n\over2};
-{{m_{N}^{2}}\over{4\epsilon_{0}^{2}}}x^{2}\right)\nonumber\\[0.2cm]
& & -\left.\frac{m_N^2}{4\epsilon_0^2}\frac{\Gamma({9\over2})}{\Gamma(7)}
\int_0^1\,dx\,g(x,\epsilon_0^2)\,{}_3F_2\left(1,{9\over4},{11\over4};
{7\over2},4;-{{m_{N}^{2}}\over{4\epsilon_{0}^{2}}}x^{2}\right)\right].
\label{Ampl3}
\end{eqnarray}
The quantities $r_{0}=4/(3\alpha_{s}m_c)$ and 
$\epsilon_{0}=m_{c}(3\alpha_{s}/4)^2 $ correspond to the 'Bohr' radius and 
the 'Rydberg' energy of the lowest $c\bar{c}$ bound state $J/\psi$, with
$m_c$ for the mass of the charm quark. The gluon distribution
$g(x,\epsilon_0^2)$ is renormalized at the quarkonium binding energy scale 
$\epsilon_0$. Eq.\ (\ref{Ampl3}) includes target mass corrections
\cite{KSSZ}; neglecting these would lead back to the
form used in \cite{KS94}. This form encounters problems in calculating 
the real part and hence would here give an incorrect threshold behavior.

\begin{figure}[htb]
\epsfysize=10cm
\centerline{\epsffile{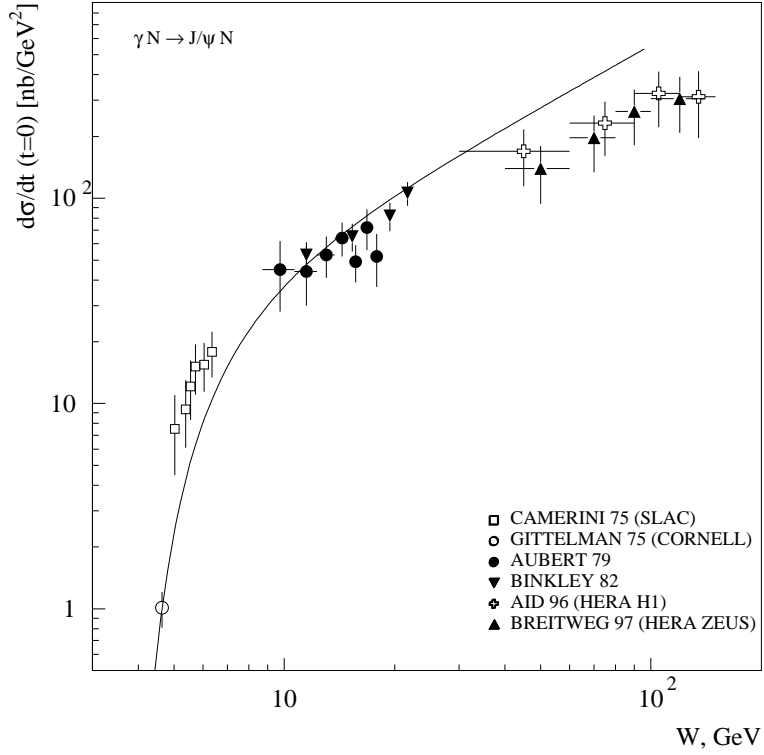}}
\caption{The same as in Fig.1, but with the curves 
obtained using the PDF MRS H.}
\end{figure}

Taking into account the obvious analytical properties of the amplitude
(\ref{Ampl3}), one now relates the physical to the unphysical regions in
$\lambda$; together with the optical theorem (\ref{OptTheor})
this leads to sum rules for the $J/\psi-N$ cross section,
\begin{eqnarray}
& & \int_0^1\,dy\,y^{n-2}(1-y^2)^{1/2}
\sigma_{\psi N}^{tot}(m_N/y) =\nonumber\\[0.2cm]
& & I(n)\int_0^1\,dx\,x^{n-2}\,
g(x,M_q^2)\,{}_3F_2\left({5\over4}+{n\over2},
{7\over4}+{n\over2},1+n;{{(5+n)}\over{2}},3+{n\over2};
-{{m_{N}^{2}}\over{4\epsilon_{0}^{2}}}x^{2}\right),
\label{SumRules}
\end{eqnarray}
where $y=m_N/\lambda$ and
$I(n)=(\pi^{3/2}/2)(32/3)^2[\Gamma(n+5/2)/\Gamma(n+5)]r_0^3\epsilon_0
(m_N/\epsilon_0)^{n-1}$.
In a first iteration, the solution of these sum rules can be written
as a convolution of the gluon distribution function and the
gluon-$J/\psi$ cross section (see \cite{PeskinBhanot}),
\begin{eqnarray}
\sigma_{N\psi}^{(0)}\left(\lambda\right)=
\frac{8\,\pi}{9}\left(32 \over 3 \right)^2\
{1 \over {\alpha_S\,m_c^2}}\
\int_{\epsilon_0/\lambda}^1\,
dx\frac{(\left(x\lambda/\epsilon_0)-1\right)^{3/2}}
{\left(x\lambda/\epsilon_0\right)^5}\,{\ds\frac{g(x,M_q^{2})}{x}},
\label{PartonicCS}
\end{eqnarray}
neglecting terms of order ${m_N}^2/{\epsilon_0}^2$. Including higher order
terms, the full solution can be obtained iteratively, each step providing a
contribution from the corresponding term of the hypergeometric series. 
Effectively, these target mass corrections change the $x$-variable in the 
convolution (\ref{PartonicCS}) and thus the resulting threshold behaviour.

\medskip

\begin{figure}[htb]
\vspace*{-0mm}
\epsfysize=10cm
\centerline{\epsffile{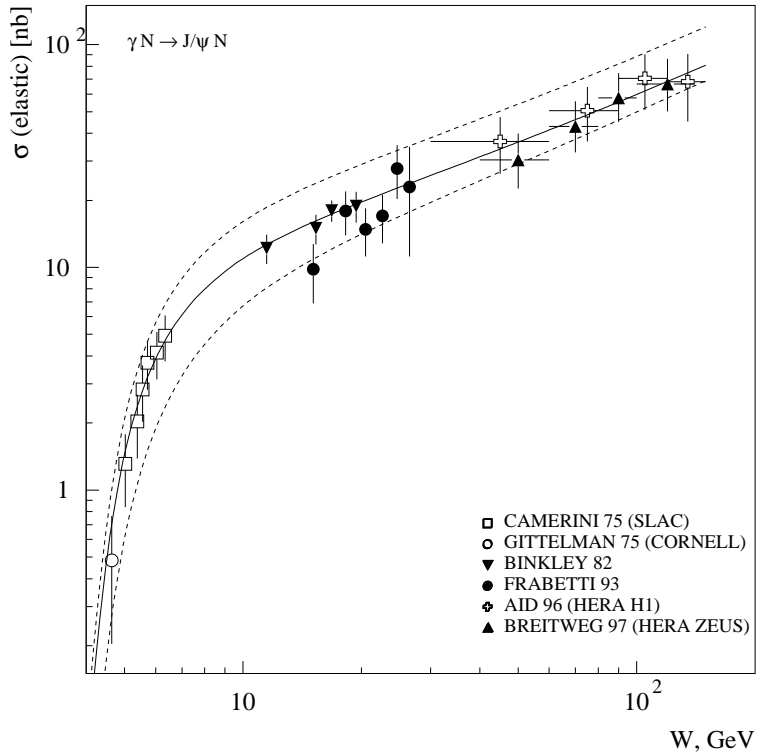}}
\caption{Data for the elastic $J/\psi$-photoproduction 
cross section, compared to the prediction obtained with the PDF MRS H.}
\end{figure}

For a given gluon distribution $g(x,\epsilon_0)$, we thus obtain
$\sigma_{\psi N}^{tot}$ and hence the imaginary part of the
$J/\psi - N$ scattering amplitude. To determine the 
forward $J/\psi$-photoproduction cross section (see Eq's. (\ref{VMD0}) and 
(\ref{VMD00})), we need the real part of the amplitude as well. The high 
energy behaviour of the corresponding amplitude makes it possible to express 
this in terms of dispersion integrals with one subtraction, performed e.g.\ at
$\lambda=0$ \cite{NSVZ78,NSVZ81}:
\begin{equation}
Re\,{\cal M}_{\psi\,N}(\lambda)={\cal M}_{\psi\,N}(0)+
\frac{2\,\lambda^2}{\pi}\int_{\lambda_0}^{\infty}\,
\frac{d\,\lambda^{'}}{\lambda^{'}}\frac{Im\,
{\cal M}_{\psi\,N}(\lambda')}{\lambda^{'2}-\lambda^{2}}.
\label{dispersion}
\end{equation}
The subtraction constant ${\cal M}_{\psi\,N}(0)$, which is needed
to estimate the behaviour of the real part of the amplitude near 
threshold is obtained by calculating the corresponding limit 
($\lambda \rightarrow M_{\psi})$ in Eq.\ (\ref{Ampl3}). This is a
self-consistent approach if we are given the analytical expression 
Eq.\ (\ref{Ampl3}) for the amplitude. A more
phenomenological and perhaps also more realistic procedure would be 
to use the low-energy theorems to estimate the subtraction
constant, following some recent work \cite {Kharz1}. 

\medskip

To complete our calculation, we have to specify the gluon distribution
function of the nucleon and fix the overall normalization in terms of
the different constants in Eq.\ (\ref{VMD1}) and related quantities. For
$g(x,m_{c}^2)$, we use two parametrizations specified in \cite{KS94}
and \cite{MRSH}. The MRS H of \cite{MRSH} parametrization, in particular,
takes into account the recent HERA results at small $x$ and thus seems
preferable for our analysis. Although all quantities in our formulae
are in a sense `physical' constants, uncertainties enter
through the c-quark mass $m_c$ and the resulting values of $\alpha_s$ 
and $\epsilon_0$; an addition, both the VMD model and the Coulombic
description of the $J/\psi$ may require corrections. 
We therefore treat the overall normalization of our 
result as an open constant, as was also done in \cite{PeskinBhanot}.

\medskip

We now want to compare our results with the data available from
low energy Cornell-SLAC \cite{SLAC} up to recent high energy HERA studies
\cite{DESY}. These experiments measure the photoproduction cross section
as function of the invariant momentum transfer $t$ in the `diffractive'
region at small $t$. One can now fit the $t$-dependence in the customary
exponential form $\exp{(-b\,t)}$ and then either extrapolate it to
the unmeasurable limit $t=0$ at which our prediction (\ref{VMD0}) holds,
or integrate over $t$ from $t_0=t_{\mbox{max}}(s)$ to infinity to obtain
the `elastic' photoproduction cross section 
$\sigma_{\gamma N\rightarrow J/\psi N}$(elastic).

\medskip

Figs.\ 1 gives the c.m.s.\ energy dependence of the forward 
differential cross section for the scaling PDF-parameterization $g(x)=
2.5(1-x)^4$ used in \cite{KS94};
we show both our complete result and the form obtained by neglecting the
real part, as in \cite{KS94}. It is seen that the inclusion of the real part 
greatly improves the threshold behaviour; the agreement is now quite good,
except for the high energy data, for which the small $x$ behaviour of the
PDF becomes important. Since the PDF form used here does not include this,
it is clear that there will be deviations at high $W$. In Fig.\ 2, we then 
show the corresponding result for the new PDF MRS H, which does include
the small $x$ results from Hera. While the qualitative agreement is 
reasonable, there are definite deviations; these would become weaker for
a less singular small $x$ form of the PDF. Moreover, there is some 
experimental discrepancy between $t$-dependence of the SLAC data compared
to other data in the same energy region; we shall return to this shortly. 
It is also not clear if a fine-tuning of $m_c$ and $\alpha_S$ would improve 
the situation.

\begin{figure}[htb]
\vspace*{-0mm}
\epsfysize=9cm
\centerline{\epsffile{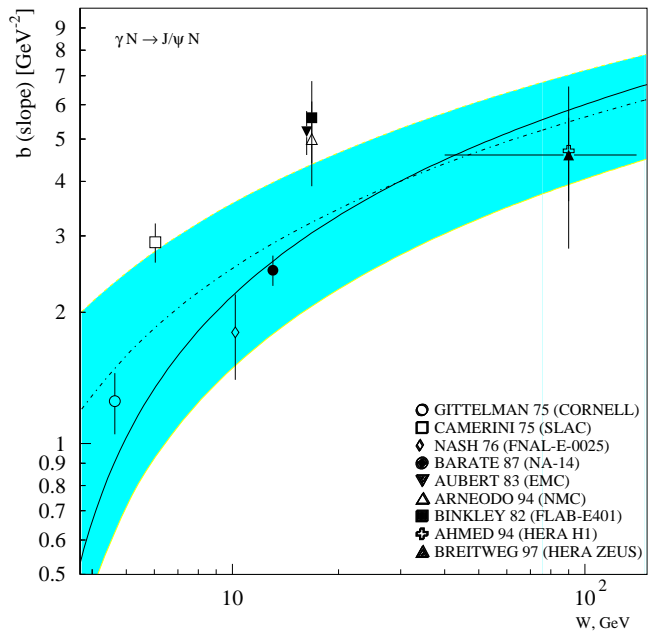}}
\caption{The energy dependence of the
slope parameter for differential elastic $J/\psi$-photoproduction.
The solid line corresponds to a best fit with $b=a_0 + a_1 \ln W^2$,
giving $a_0=-1.64 \pm 0.26$ and $a_1=0.83 \pm 0.06$;
for data references, see \cite{DESY}}
\end{figure}

In Fig.\ 3, we then show our results for the exclusive $J/\psi$
photoproduction cross section as a function of $W$, compared to relevant 
experimental data; the theoretical error range is explained below.
The curve was obtained for the MRS H parameterization of 
the PDF, using the approach mentioned above. Thus, we integrate the forward 
differential cross section in the diffractive peak up to $t_{max}$ \cite{Jung}
with the slope parameter $b$ as logarithmic function of $W$ (`diffraction 
cone shrinkage'). This integrated result is much less sensitive to the
parameter values involved and provides a good interpolation between the
high energy (small $x$) data and the near-threshold behaviour of the
cross section. 

\medskip

Finally, we want to return shortly to the slope parameter parametrisation we 
have used above. In Fig.\ 4 we show the result of fitting the form $b=a_0 
+a_1\log W^2$ 
to all available measurements; a best fit gives
$a_0=-1.64 \pm 0.26$ and $a_1=0.83 \pm 0.06$. It is seen that the SLAC 
data give
a slope parameter considerably higher than that of the other experiments
in the same energy region. This leads directly to the comparatively high
values of $d\sigma/dt(t=0)$, which are obtained by an extrapolation to $t=0$
using the measured slope parameter. -- The energy variation of $b$ indicates
the presence of contributions from `soft' as well as from a `hard' Pomeron
\cite{DL95}. Here the role of the low energy points is crucial; a restriction 
to only high energy points \cite{FKS} could lead to an energy-independent
slope as expected for a `hard' Pomeron.

\medskip

In conclusion, we note that our analysis of photoproduction confirms 
the relation between the energy dependence of the cross section and the
$x$-dependence of the gluon distribution function of the nucleon \cite{KS94}. 
This relation was derived for $J/\psi$-hadron interactions and enters
photoproduction through VMD. The success of such a consistent treatment 
of these two reactions does not support expectations about inherently
different behaviour \cite{FKS}.

\bigskip

\noindent{\bf Acknowledgements}

\medskip

It is a pleasure to thank D. Schildknecht and H. Spiessberger for stimulating
discussions.

\end{document}